\documentclass[prl, twocolumn]{revtex4-1}

\usepackage{graphicx} 
\usepackage{amssymb} 
\usepackage[usenames]{color}
\usepackage{amsmath}
\usepackage{xspace}

\begin{document}

\title{Dimensionality selection in a molecule-based magnet}

\author{Paul A. Goddard,$^{1}$ Jamie L. Manson,$^{2}$ John Singleton,$^{3}$
Isabel Franke,$^{1}$ Tom Lancaster,$^{1*}$ Andrew J. Steele,$^{1}$
Stephen J. Blundell,$^{1}$ Christopher Baines,$^{4}$ Francis L. Pratt,$^{5}$
Ross D. McDonald,$^{3}$ Oscar E. Ayala-Valenzuela,$^{3}$ Jordan F. Corbey,$^{2}$
Heather I. Southerland,$^{2}$ Pinaki Sengupta,$^{6}$ and John A. Schlueter$^{7}$}

\affiliation{$^{1}$University of Oxford, Department of Physics, Clarendon Laboratory, Parks Road, Oxford OX1~3PU, U.K.\\
$^{2}$Department of Chemistry and Biochemistry, Eastern Washington University, Cheney, WA 99004, USA\\
$^{3}$National High Magnetic Field Laboratory, Los Alamos National Laboratory, MS-E536, Los Alamos, NM 87545, USA\\
$^{4}$Paul Scherrer Institut, Laboratory for Muon-Spin Spectroscopy, CH-5232 Villigen PSI, Switzerland\\
$^{5}$ISIS Facility, STFC Rutherford Appleton Laboratory, Didcot, Oxfordshire, OX11 0QX\\
$^{6}$School of Physical and Mathematical Sciences, Nanyang Technological University, 21 Nanyang Link, Singapore 637371\\
$^{7}$Materials Science Division, Argonne National Laboratory, Argonne, IL 60439, USA}

\begin{abstract}
Gaining control of the building blocks of magnetic materials and thereby achieving particular characteristics will make possible the design and growth of bespoke magnetic devices. While progress in the synthesis of molecular materials, and especially coordination polymers, represents a significant step towards this goal, the ability to tune the magnetic interactions within a particular framework remains in its infancy. Here we demonstrate a chemical method which achieves dimensionality selection via preferential inhibition of the magnetic exchange in an $S=1/2$ antiferromagnet along one crystal direction, switching the system from being quasi-two- to quasi-one-dimensional while effectively maintaining the nearest-neighbour coupling strength.
\end{abstract}

%%%%%%%%%%%%%%%%% END OF PREAMBLE %%%%%%%%%%%%%%%%

\maketitle

Coordination polymers are self-organising materials consisting of arrays of metal ions linked via molecular ligands, with non-coordinated counterions supplying charge neutrality. The choice of initial components permits a high level of control over the final product, enabling many different polymeric architectures to be obtained~\cite{batten_book}. These materials provide a route to successful crystal engineering, and a number of functionalities are being actively studied, including gas storage~\cite{farha_10,bureekaew_09,rosi_03}, optoelectronic~\cite{kuchison_10,katz_09}, ferroelectric~\cite{sengupta_10,okubo_05} and magnetic properties~\cite{ohkoshi_11,halder_11,her_10,fishman_10,musfeldt_09,mansonjacs_09}. 

Although it is now possible to generate an assortment of disparate magnetic lattices using this method~\cite{miller_11,blundell_04}, true control of magnetic exchange interactions implies an ability to adjust selected parameters while keeping others constant. To this end, a series of coordination polymers based on Cu(II) ions bridged by pyrazine (C$_4$H$_4$N$_2$) molecules have proven to be highly versatile. In these systems it has been shown that it is possible to alter significantly the primary exchange energies via adjustment of the ligands~\cite{butcher_08} and the counterions~\cite{goddardnjp_08,woodward_07}, or fine-tune the exchange by a few percent via isotopic substitution~\cite{goddardprb_08}, all the while maintaining the same basic metal\textendash pyrazine network. In this paper we demonstrate the power of this strategy by chemically engineering a reduction in the dimensionality of a magnetic system. After first designing a material based on coordinated planes of Cu(II), we adapt the recipe such that the ligand bridges are broken along a specific crystal direction, resulting in a chain-like compound. Because the ligand mediating the magnetic interactions in both cases is unchanged, the nearest-neighbour exchange energies of the two materials are found to be equal to each other to within 5\%. The difference in numbers of nearest-neighbours, however, means that the strength of the combined exchange interactions acting on each magnetic ion in the quasi-two-dimensional material is twice that of its quasi-one-dimensional cousin.

\begin{figure*}[t]
\centering
\includegraphics[width=8.9cm]{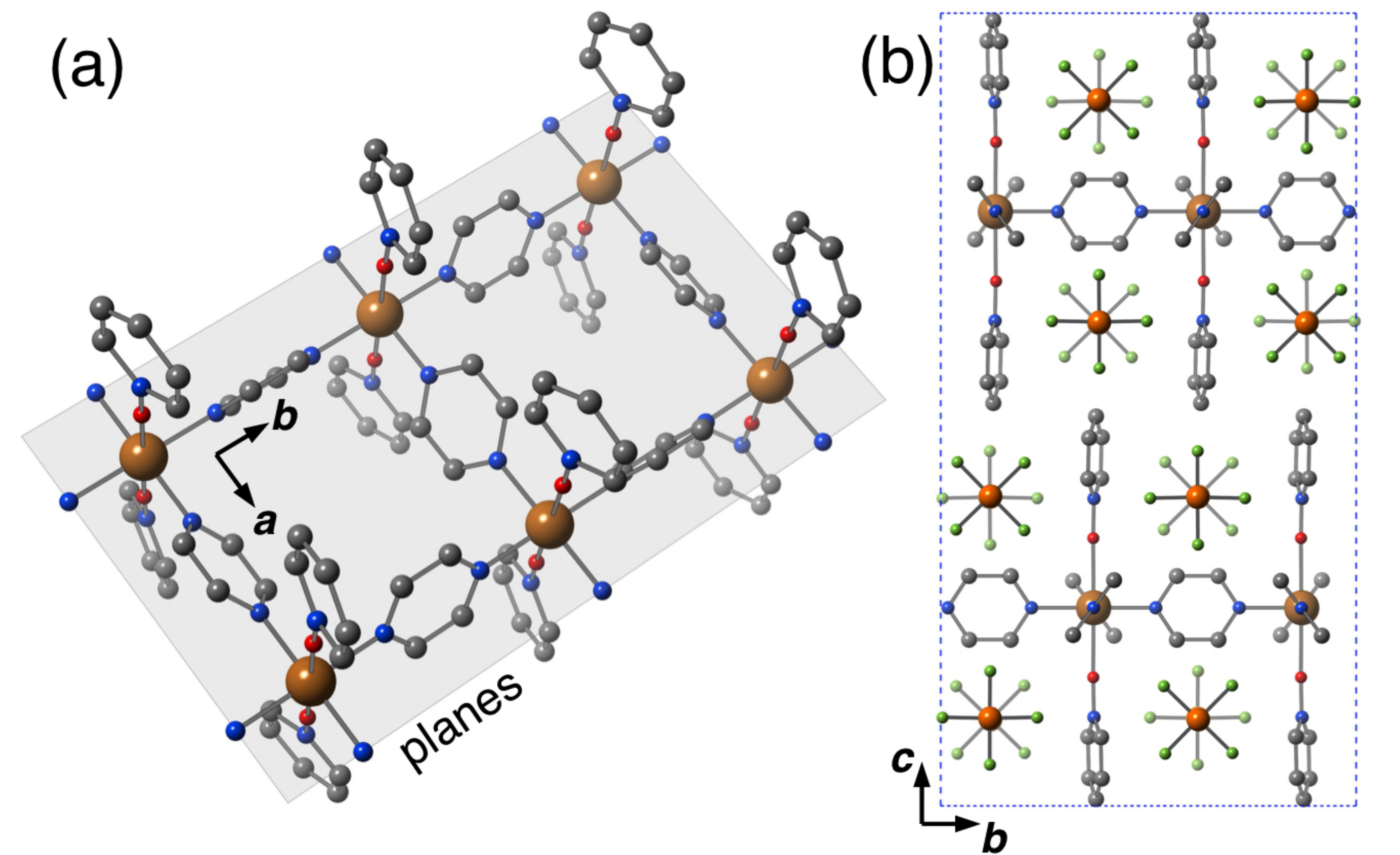}
\includegraphics[width=8.9cm]{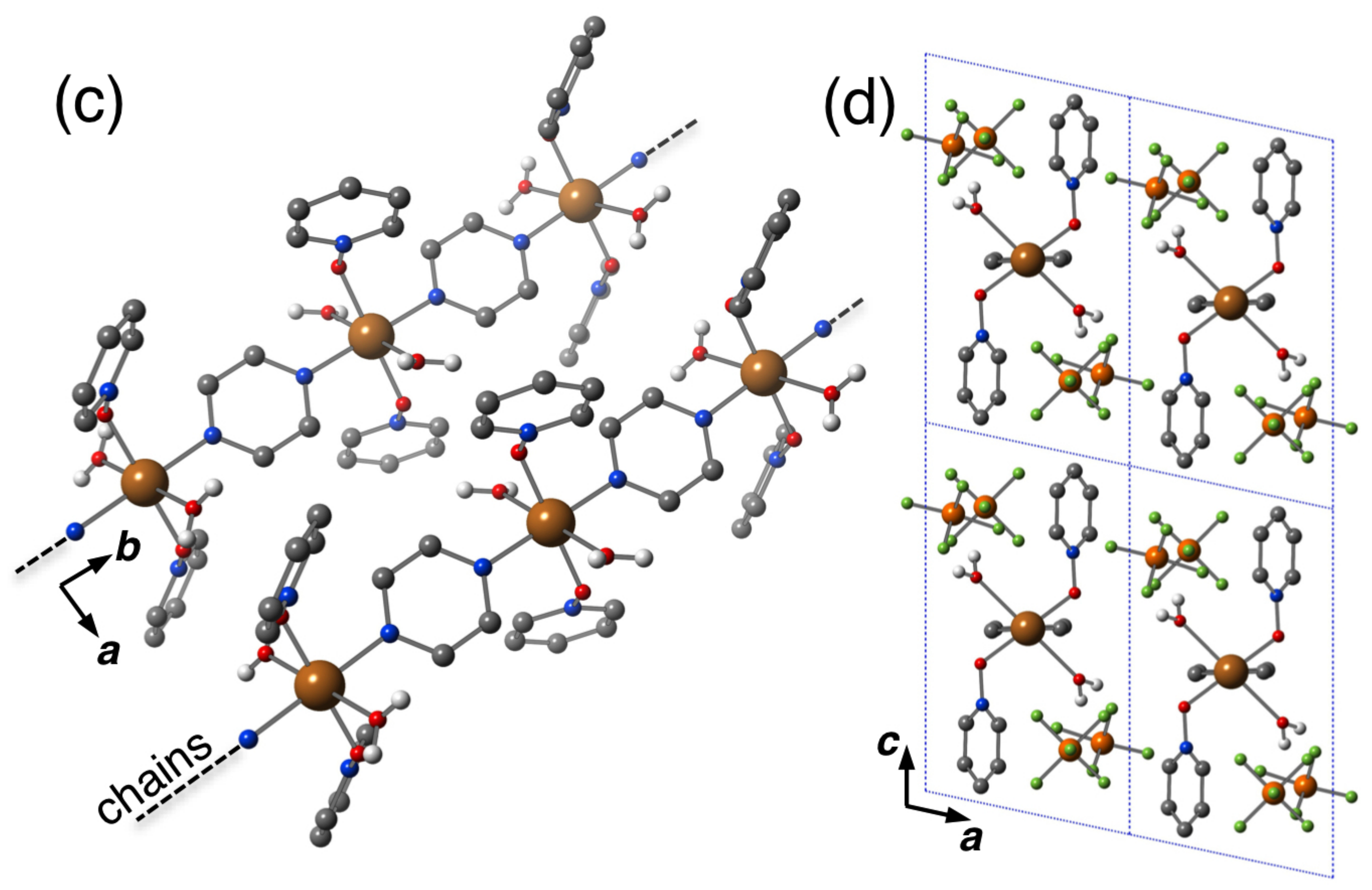}
\caption{ (a) View of the crystal structure of the planar material [Cu(pyz)$_2$(pyO)$_2$](PF$_6$)$_2$  determined using x-ray diffraction showing the 2D Cu\textemdash pyrazine network in the $ab$-plane. (b) a projection of the same structure along the ${\bf a}$-axis highlighting the shift between adjacent Cu\textemdash pyrazine layers and the arrangement of the PF$_6$ counterions. (c) Crystal structure of the chain-like material [Cu(pyz)(pyO)$_2$(H$_2$O)$_2$](PF$_6$)$_2$ showing the 1D Cu\textemdash pyrazine chains in the $ab$-plane, and (d) a projection along the chains showing the arrangement of the pyO and H$_2$O ligands and the PF$_6$ counterions. The dashed lines in (c) and (d) enclose one unit cell. Cu = brown, C = grey, N = blue, O = red, P = orange, F = green. Hydrogens other than those in the water molecules have been omitted for clarity. }
\label{fig1}
\end{figure*}

Figs.~\ref{fig1}(a) and (b) show the crystal structure of orthorhombic [Cu(pyz)$_2$(pyO)$_2$](PF$_6$)$_2$ (where pyz = pyrazine and pyO = pyridine-$N$-oxide, C$_5$H$_5$NO) determined using single-crystal x-ray diffraction~\cite{suppinfo}. $S=1/2$ Cu ions are linked by pyz molecules into nearly square planar arrays, with perpendicular non-bridging pyO ligands keeping the planes well-separated. Because of the separation, as well as the staggered arrangement of adjacent planes shown in Fig.~\ref{fig1}(b), magnetic exchange energies are likely to be very small along the ${\bf c}$-direction. In contrast, Cu\textemdash pyz\textemdash Cu bridges are known to be good mediators of antiferromagnetic superexchange~\cite{mansoncc_06, darriet_79} and so the magnetic properties of this material are expected to be quasi-two-dimensional. This is confirmed by the magnetic measurements described below. Sample synthesis involves mixing together of the molecular components in a solution of water and ethanol. Intermolecular self-organisation means that only a small amount of intervention is subsequently required. To achieve the desired planar structure the pyz and pyO molecules were added in a 3:1 ratio, previous experience suggesting that to account for the potential for pyO to substitute for pyz, the ligands must initially be in a proportion different to that found in the final product. In order to create a similar sample, but one based on Cu\textemdash pyz chains rather than planes, we reduce the pyz:pyO ratio to 2:1 and proceed with the synthesis in a similar way. The resulting material has the composition [Cu(pyz)(pyO)$_2$(H$_2$O)$_2$](PF$_6$)$_2$ and the structure is shown in Figs.~\ref{fig1}(c) and (d). Here the pyz ligands link Cu ions along the ${\bf b}$-axis only, the other ligands being non-bridging pyO and water molecules. The alteration in composition has the effect of changing the symmetry of the crystal from orthorhombic to monoclinic and reducing the number of formula units in the unit cell, but, most importantly for the magnetic exchange, the Cu\textemdash pyz\textemdash Cu linkages along the ${\bf a}$-axis are removed without altering the ${\bf b}$-axis Cu--Cu separation by more than a fraction of a percent ($6.914\pm0.001$~\AA  ~for the planar material and $6.851\pm0.001$~\AA  ~for the chain compound).

\begin{figure}[t!]
\centering
\includegraphics[width=8.0cm]{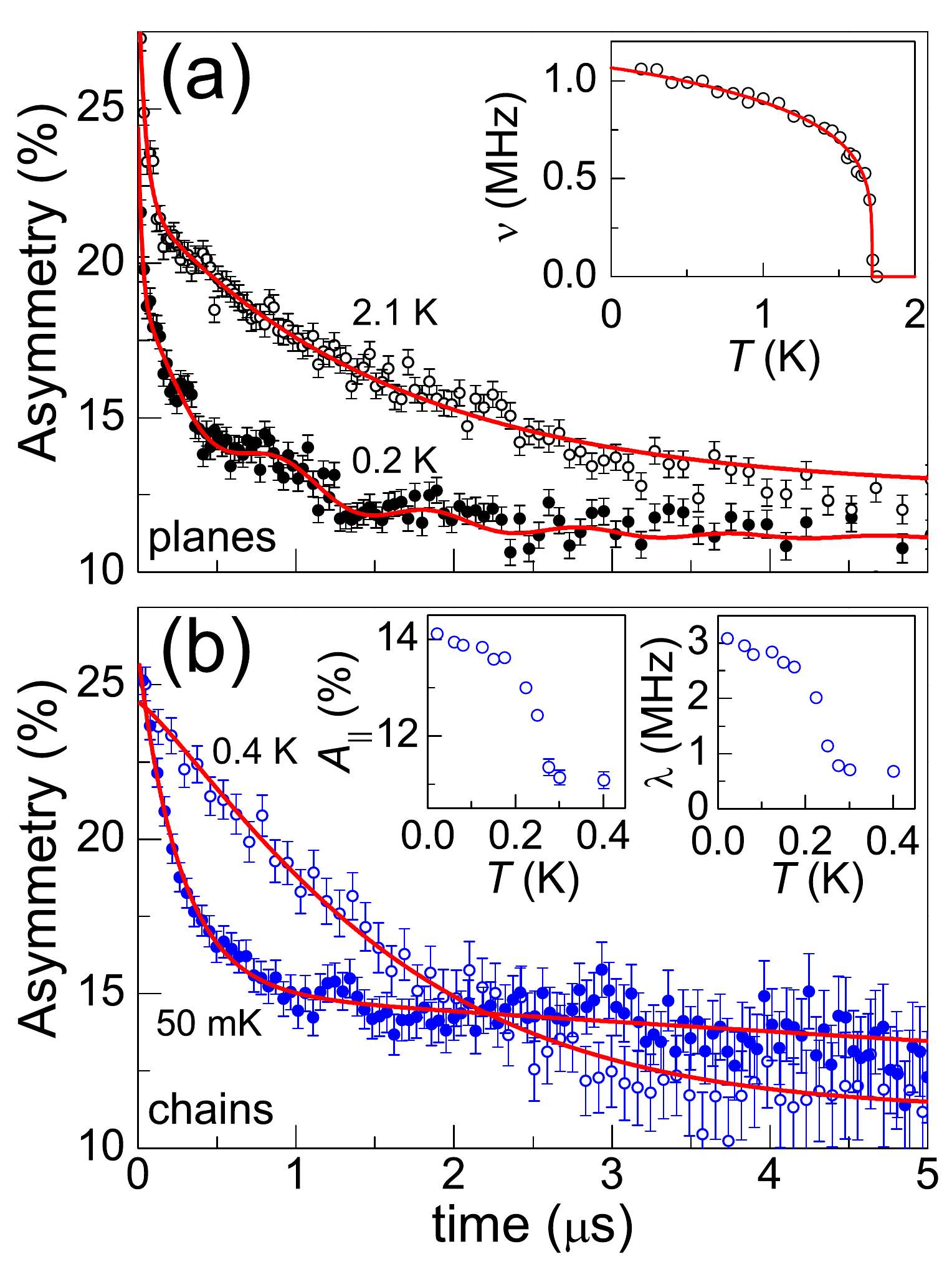}
\caption{(a) Example muon-spin relaxation ($\mu^{+}$SR) spectra measured on planar
[Cu(pyz)$_{2}$(pyO)$_{2}$](PF$_{6}$)$_{2}$. Data at different temperatures are offset for clarity. Inset: Evolution of the precession frequency $\nu$ with temperature. Long-range magnetic order is observed below 1.71~K.  (b) $\mu^{+}$SR spectra measured on chain-like [Cu(pyz)(pyO)$_{2}$(H$_{2}$O)$_{2}$](PF$_{6}$)$_{2}$.
Insets: The evolution of the amplitude $A_{\parallel}$ (left) and
relaxation rate $\lambda$ (right) with temperature indicates the onset of long-range magnetic order below about 0.27~K. Red lines are fits to functions described in the text and~\cite{suppinfo}.}
\label{muon}
\end{figure}

Recent heat capacity measurements on the planar material in zero magnetic field see no evidence of a magnetic transition down to the lowest temperatures measured~\cite{kohama_11}. However, thermodynamic probes are known to be less sensitive to transitions driven by interplanar couplings~\cite{sengupta_03} while local probes such as muon-spin relaxation ($\mu^{+}$SR) are much more effective at determining the antiferromagnetic transition temperature, $T_{\rm N}$~\cite{mansoncc_06}.  Our $\mu^+$SR data on this compound, shown in Fig.~\ref{muon}(a), exhibits a clear precession signal which develops below $T_{\mathrm{N}}=1.71\pm0.02$~K \cite{steele11}, demonstrating long-range magnetic order throughout the bulk of the sample.  In contrast, the $\mu^{+}$SR data for the chain-like compound exhibit no resolvable oscillations, see Fig.~\ref{muon}(b), probably due to a smaller ordered moment, but can be fitted to the expression $A(t) = A_{0}e^{-\lambda t} + A_{\parallel}$, where $A_{\parallel}$ represents the non-relaxing part of the signal.  Both $A_{\parallel}$ and $\lambda$ rise markedly below $T=0.26$~K due to crossover from a regime in which the relaxation is dominated by dynamic magnetic fluctuations to one dominated by quasistatic magnetic order, see Fig.~\ref{muon}(b) insets. From these fits we estimate $T_{\mathrm{N}} = 0.27\pm0.01$~K for this material~\cite{suppinfo}.

\begin{figure}[!t]
\centering
\includegraphics[width=8.5cm]{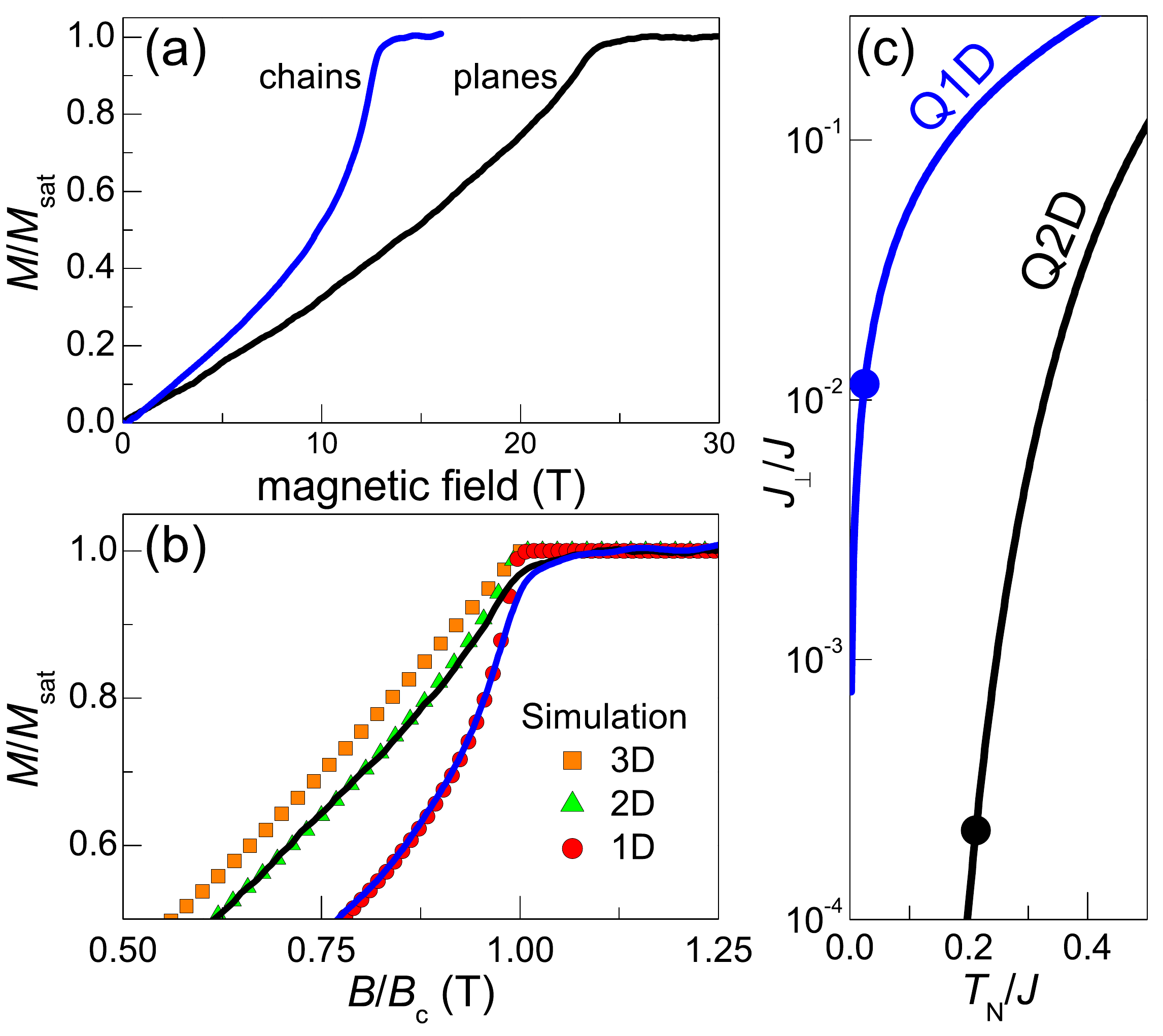}
\includegraphics[width=8.5cm]{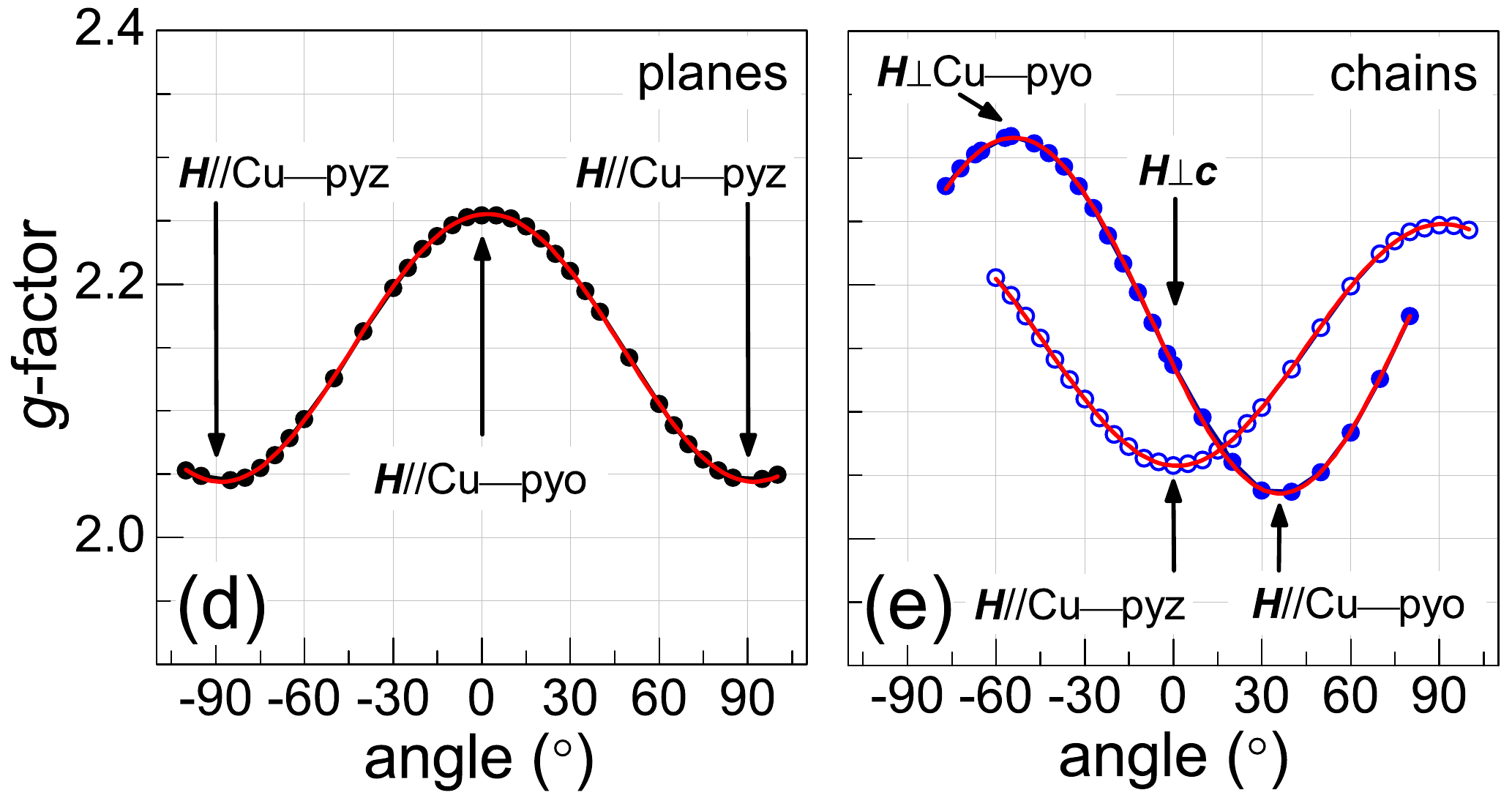}
\caption{(a) Normalized pulsed-field magnetization of planar
[Cu(pyz)$_{2}$(pyO)$_{2}$](PF$_{6}$)$_{2}$  at $T=1.5$~K, and chain-like [Cu(pyz)(pyO)$_{2}$(H$_{2}$O)$_{2}$](PF$_{6}$)$_{2}$ at $T=0.5$~K. (b) The results of quantum Monte Carlo (QMC) simulations of the low temperature magnetization for 3D ($J_{\perp}/J  = 1$), 2D ($J_{\perp}/J  = 0$), and 1D ($J_{\perp}/J  = 0.02$) antiferromagnets. The lines are the pulsed-field data scaled by the saturation field ($B_{\rm c}$). (c) The relation between the exchange anisotropy and the ratio of critical temperature and primary exchange energy in Q1D and Q2D antiferromagnets deduced from QMC simulations~\cite{yasuda_05}. The circles indicate the materials reported here. (d) Anisotropy of the $g$-factor in the planar and (e) chain-like compounds measured using ESR at 10~K and 1.5K, respectively. Red lines are fits to $g(\theta)=(g_{xy}^2\sin^2\theta+g_z^2\cos^2\theta)^{\frac{1}{2}}$.}
\label{pulsed}
\end{figure}

The type of magnetic order displayed by the two compounds can be deduced from their low-temperature magnetization (see Fig.~\ref{pulsed}(a)). The form of our pulsed-field magnetization data up to saturation is in keeping with that expected for low-dimensional antiferromagnets. The slightly concave curve exhibited by the planar material is typical of quasi-two-dimensional (Q2D) antiferromagnetic interactions~\cite{goddardnjp_08}, while the more extreme curvature shown by data from the chain-like sample is indicative of a quasi-one-dimensional (Q1D) magnetic lattice, where, for an ideal system, ${\rm d}M/{\rm d}B$ is known to diverge at the saturation field~\cite{bonnerfisher}. To support these observations we compare the data with the results of low-temperature quantum Monte Carlo (QMC) simulations based on the Hamiltonian
\begin{equation}
{\cal H} =J\sum_{\langle i,j \rangle_\parallel}{\bf S}_i\cdot {\bf S}_j +
J_\perp\sum_{\langle i,j \rangle_\perp}{\bf S}_i\cdot {\bf S}_j 
-g\mu_{\rm B}B\sum_iS_i^z.
\label{ham}
\end{equation}
Here, for a Q1D (or Q2D) system, $J$ is the strength of the exchange coupling within the magnetic chains (planes), $J_\perp$ is the coupling between chains (planes), and the first and second summations refer to summing over unique pairs of nearest neighbours parallel and perpendicular to the chain (plane), respectively. Comparisons of theory and data are shown in Fig. 3(b). Theoretical magnetisation curves were calculated in finite steps of the $J_{\perp}/J$ parameter for both Q1D and Q2D magnetic lattices~\cite{goddardnjp_08,suppinfo}. For both materials curves corresponding to the quoted values of $J_{\perp}/J$ gave the best matches with experimental data; the Q2D curve with $J_{\perp}/J  = 0.00$ for the planar sample, and the Q1D curve with $J_{\perp}/J  = 0.02$ for the chain-like material.  The predicted curve for a three-dimensional antiferromagnet is also shown for contrast. The deviation of the data from the simulations close to the saturation field is likely due to the finite temperatures at which the experiments were performed. 

The saturation field ($B_{\rm c}$) can be extracted from the pulsed-field data and is found to be $23.7\pm0.8$~T for the planar material and $12.8\pm0.4$~T for the chain-like material. At saturation, the Hamiltonian above implies that $g\mu_{\rm B}B_{\rm c}=nJ+n_\perp J_\perp$, where, for each spin, $n$ is the number of exchange bonds of interaction strength $J$ ($n=4$ for Q2D, 2 for Q1D), and $n_\perp$ is the number of exchange bonds of interaction strength $J_\perp$ ($n_\perp=2$ for Q2D, 4 for Q1D)~\cite{goddardnjp_08,goddardprb_08,steele11}. From this, by assuming that $J_\perp/J\ll1$ and using the appropriate values for the $g$-factor determined from electron-spin resonance, we estimate the primary exchange couplings, $J$, to be $8.1\pm0.3$~K for the planar compound and $8.8\pm0.2$~K for the chain-like compound.

Another estimate of the magnetic dimensionality comes from the temperature dependence of the low-field magnetic susceptibility. Fits of such data for both materials results in estimates of $J$ that are in accord with those derived above from pulsed-field magnetization data~\cite{suppinfo}.

The relative sizes of $T_{\rm N}$ and $J$ are indicative of the anisotropy of the exchange interactions in a low-dimensional magnetic system. Using quantum Monte Carlo (QMC) calculations Yasuda {\it et al.}~\cite{yasuda_05} developed empirical relations between these values for $S = 1/2$ Heisenberg antiferromagnets, the results of which are shown in Fig.~\ref{pulsed}(c). For compounds considered here the exchange anisotropies are found to be $|J_\perp/J| \sim 10^{-4}$ in the planar material and $10^{-2}$ in the chain-like material. These values are in keeping with the comparison between data and simulation shown in Fig.~\ref{pulsed}(b).

The magnetic lattice in a low-dimensional system is not always obvious from an inspection of the crystal structure~\cite{mansoncm_08, butcher_08}. Verification of the equivalence of the magnetic and structural planes and chains in our materials comes from the angle-dependence of the $g$-factor as determined by electron-spin resonance. To first approximation, the Cu(II) ions in the planar material have a local octahedral symmetry with a tetragonal distortion along the ${\bf c}$-axis. In such situations, the unpaired spin in a $d^9$~configuration is expected to occupy the $d_{x^2-y^2}$ orbital~\cite{abragambook}, and the $g$-factor parallel to the distorted $z$-axis takes a larger value than those in the $xy$-plane (see discussion in Ref.~\onlinecite{mansoncm_08}). Accordingly, the data in Fig.~\ref{pulsed}(d) show that the $g$-factor is lowest when the excitation field is applied in the $ab$-plane, implying that the direction of highest electronic orbital overlap is the Cu\textemdash pyz directions, with the strong Q2D exchange interactions being mediated via the molecular orbitals of the pyrazine. For the chain-like material, the distortion is more complicated due to the lower crystal symmetry. Nevertheless, using two rotations, the smallest values of the $g$-factor are found to be along the Cu\textemdash pyz and Cu\textemdash pyO bonds, as shown in Fig.~\ref{pulsed}(e), implying that again the magnetic orbital is $d_{x^2-y^2}$. This allows for the possibility of good exchange coupling along those bonds, but as the pyO molecules are non-bridging ligands, the Q1D magnetism must be mediated along the Cu\textemdash pyz chains. 

Taken together, these experimental observations paint a complete picture of the two closely-related magnetic systems. From the point of view of magnetic superexchange, the chain direction in the Q1D material looks very similar to the two Cu\textemdash pyz directions in the Q2D material and so, despite the compositional differences, the primary nearest-neighbour exchange energies remain largely unaltered. At the same time the critical field in the Q2D material is approximately double that for the Q1D compound because it has twice the number of nearest neighbours. The Q2D compound is strongly anisotropic, comparable to the most 2D materials yet identified~\cite{steele11}. As suggested above, this is likely due to the disconnect and staggering that occurs between successive planes. The extreme anisotropy explains why the zero-field heat capacity was not sensitive to the antiferromagnetic transition observed using $\mu^{+}$SR. In the Q1D material the anisotropy is less pronounced (even though the $T_{\rm N}/J$ ratio is smaller) because the chains are not staggered, there are twice as many next-nearest neighbours, and the shortest distance between chains (along the ${\bf a}$-axis) is approximately half the interlayer separation in the Q2D material.

These materials showcase the ability to take deliberate control over the magnetic properties of polymeric systems. The self-organisation of the coordination polymers enables them to spontaneously form crystalline lattices whose structure can be anticipated with a high level of predictability. It is this predictive power, together with the ability to choose the starting ingredients and the knowledge of the exchange efficiency of various ligands accrued over the past few decades, that permits the preselection of exchange anisotropy. In the Cu\textemdash pyz systems we have demonstrated the ability to make this preselection without significantly perturbing the magnitude of the primary interaction strengths, while previous studies of similar materials have highlighted the capacity to tune the exchange couplings without changing the overall dimensionality. Thus these compounds represent a promising approach to magnetic crystal engineering, and in particular raise the possibility of generating systems that exhibit higher ordering temperatures and other cooperative phenomena.

This project was supported by the EPSRC (UK), the NSF and DoE (US), and the European Commission. We thank Peter Baker for technical assistance. PAG would like to thank Keola Wierschem for useful discussions.

\section*{Supplemental material accompanying\\
Dimensionality selection in a molecule-based magnet}

\author{Paul A. Goddard,$^{1}$ Jamie L. Manson,$^{2}$ John Singleton,$^{3}$
Isabel Franke,$^{1}$ Tom Lancaster,$^{1}$ Andrew J. Steele,$^{1}$
Stephen J. Blundell,$^{1}$ Christopher Baines,$^{4}$ Francis L. Pratt,$^{5}$
Ross D. McDonald,$^{3}$ Oscar E. Ayala-Valenzuela,$^{3}$ Jordan F. Corbey,$^{2}$
Heather I. Southerland,$^{2}$ Pinaki Sengupta,$^{6}$ and John A. Schlueter$^{7}$}

\affiliation{$^{1}$University of Oxford, Department of Physics, Clarendon Laboratory, Parks Road, Oxford OX1~3PU, U.K.\\
$^{2}$Department of Chemistry and Biochemistry, Eastern Washington University, Cheney, WA 99004, USA\\
$^{3}$National High Magnetic Field Laboratory, Los Alamos National Laboratory, MS-E536, Los Alamos, NM 87545, USA\\
$^{4}$Paul Scherrer Institut, Laboratory for Muon-Spin Spectroscopy, CH-5232 Villigen PSI, Switzerland\\
$^{5}$ISIS Facility, STFC Rutherford Appleton Laboratory, Didcot, Oxfordshire, OX11 0QX\\
$^{6}$School of Physical and Mathematical Sciences, Nanyang Technological University, 21 Nanyang Link, Singapore 637371\\
$^{7}$Materials Science Division, Argonne National Laboratory, Argonne, IL 60439, USA}

\maketitle

\noindent{\bf Synthesis of planar [Cu(pyz)$_{2}$(pyO)$_{2}$](PF$_{6}$)$_{2}$}. In the minimum amount of water, CuCl$_2\cdot$2H$_2$O (0.502 g, 2.95 mmol) and AgPF$_6$ (1.489 g, 5.89 mmol) were mixed together to yield a white precipitate of AgCl and a pale blue solution containing the ``Cu(PF$_6$)$_2$'' product. To recover the blue solution, the AgCl solid was removed by careful vacuum filtration. In a separate solution, pyrazine and pyridine-$N$-oxide in a 3:1 ratio were dissolved in a 1:1 mixture of H$_2$O and ethanol to give a colourless solution. To this latter solution was added the  ``Cu(PF$_6$)$_2$'' solution. Upon slow mixing of the chemical reagents, a small amount of pale blue powder formed, but was removed by vacuum filtration to afford a blue solution. Slow evaporation of this solution yielded a large mass of deep green square plates. 

\vspace{0.2cm}
\noindent{\bf Synthesis of the chain-like compound [Cu(pyz)(pyO)$_{2}$(H$_{2}$O)$_{2}$](PF$_{6}$)$_{2}$}. Using a reaction scheme similar to that above, the ratio of of pyrazine to pyridine-$N$-oxide was decreased from 3:1 to 2:1. Mixing of the chemical reagents immediately afforded a bright turquoise blue solution. The solution was covered with a piece of perforated Al-foil to allow slow solvent evaporation. Upon standing at room temperature for several weeks, small green blocks were deposited on the bottom of the beaker and subsequently isolated via vacuum filtration.

\begin{table}[!ht]
 \renewcommand\thetable{S1}
\centering
\begin{tabular}{|l|c|r|}
\hline
 & Q2D & Q1D \\
\hline
space group & $Cmca$ & $P2_1/m$\\
${\bf a}$ (\AA) & $13.725\pm0.002~$ & $6.948\pm0.001$ \\
${\bf b}$ (\AA) & $13.828\pm0.002~$ & $13.703\pm0.001$\\
${\bf c}$ (\AA) & $26.377\pm0.003~$ & $12.293\pm0.001$ \\
$\alpha$ ($^\circ$) & 90 & 90 \\
$\beta$ ($^\circ$) & 90 & $102.007\pm0.001$ \\
$\gamma$ ($^\circ$) & 90 & 90 \\
Cu--Cu $\parallel {\bf a}$ (\AA) & $6.863\pm0.001$  & $6.948\pm0.001$ \\
Cu--Cu $\parallel {\bf b}$ (\AA) & $6.914\pm0.001$  & $6.851\pm0.001$ \\
Cu--Cu $\parallel {\bf c}$ (\AA) & $13.189\pm0.002$* & $12.293\pm0.001$ \\
$m_{\rm mol}$ (g)& 703.86 & 659.80\\
\hline
\end{tabular}
\caption{Room temperature lattice parameters determined from x-ray crystallography of [Cu(pyz)$_2$(pyO)$_2$](PF$_6$)$_2$ (Q2D) and [Cu(pyz)(pyO)$_2$(H$_2$O)$_2$](PF$_6$)$_2$ (Q1D). *Note that this value is the perpendicular distance between Cu ions.}
\label{tablestruc}
\end{table}

\begin{table}[h]
\centering
 \renewcommand\thetable{S2}
\begin{tabular}{|lc|c|r|}
\hline
&& Q2D & Q1D \\
\hline
Cu1--N8 & (\AA) & $2.0452\pm0.0019$ & $2.044\pm0.004$\\
Cu1--N11 & (\AA) & $2.0646\pm0.0024$ & --\\
Cu1--N14 & (\AA) & $2.0690\pm0.0024$ & --\\
Cu1--O1 & (\AA) & $2.3164\pm0.0019$ & $1.969\pm0.003$\\
Cu1--O2 & (\AA) & -- & $2.420\pm0.005$\\
\hline
\end{tabular}
\caption{Coordination sphere bond lengths of \mbox{[Cu(pyz)$_2$(pyO)$_2$](PF$_6$)$_2$} and [Cu(pyz)(pyO)$_2$(H$_2$O)$_2$](PF$_6$)$_2$ (Q2D and Q1D, respectively) determined from room temperature x-ray crystallography. Atom numbering schemes can be found in Supplementary Figures S1 and S2.}
\label{tablecoord}
\end{table}

\begin{figure}[!t]
 \renewcommand\thefigure{S1}
\centering
\includegraphics[width=8.5cm]{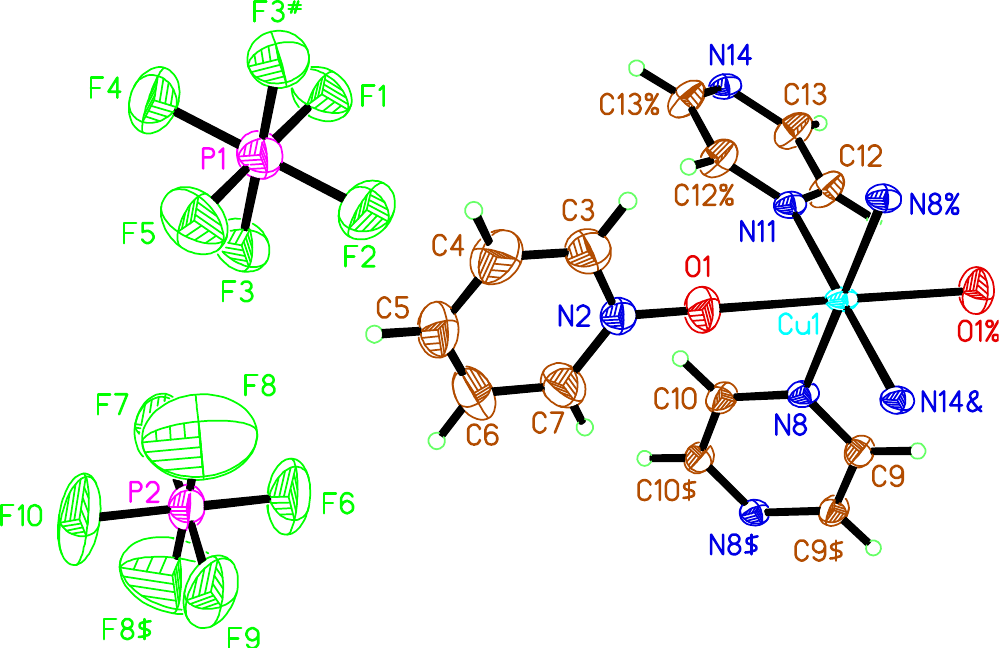}
\caption{Atom labelling scheme for the asymmetric unit of planar [Cu(pyz)$_2$(pyO)$_2$](PF$_6$)$_2$. Thermal ellipsoids are drawn at the 50\% probability level. Symmetry transformations used to generate equivalent atoms are \#: -x, y, z; \$: -x, y+1, z; \%: -x+1/2, y, -z+1/2; \&: x, y+1/2, -z+1/2.} 
\label{therm2d}

 \renewcommand\thefigure{S2}
\centering
\includegraphics[width=8.5cm]{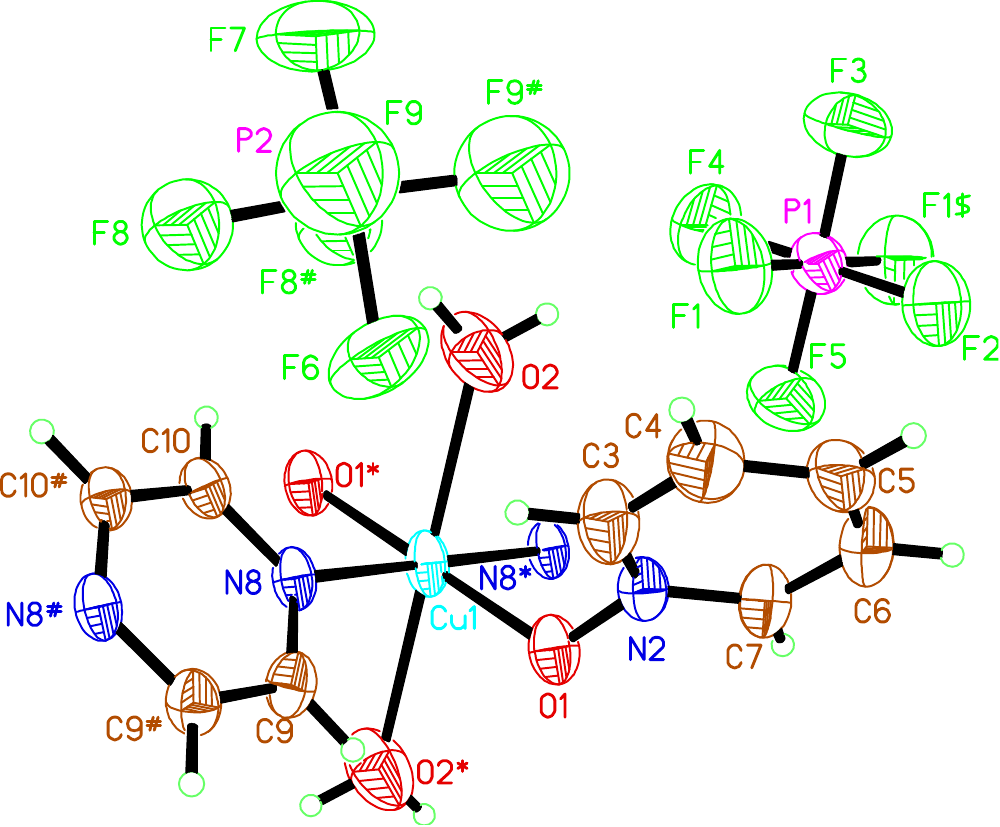}
\caption{Atom labelling scheme for the asymmetric unit of [Cu(pyz)(pyO)$_2$(H$_2$O)$_2$](PF$_6$)$_2$. Thermal ellipsoids are drawn at the 50\% probability level. Only one orientation of the disordered PF$_6$ counterion associated with P2 is shown. Symmetry transformations used to generate equivalent atoms are \#: x, -y+1/2, z; \$: x, -y+3/2, z; *: -x+1, -y+1, -z+1.} 
\label{therm2d}
\end{figure}

\vspace{0.2cm}
\noindent{\bf Structural determination.} Crystals of each compound were attached to a glass fibre and data were collected at 297(2) K using a Bruker/Siemens SMART APEX instrument (Mo K$_\alpha$ radiation, $\lambda$ = 0.71073\AA). Data were measured using omega scans of $0.3^\circ$ per frame for 10 seconds, and a full sphere of data was collected. Data were analysed and cell parameters retrieved using the supplied software. The data for \mbox{[Cu(pyz)(pyO)$_2$(H$_2$O)$_2$](PF$_6$)$_2$} were found to be rotationally twinned and were deconvolved using CELL\_NOW~\cite{cellnow} software to give a refined twinning ratio of $0.04\pm0.01$. Absorption corrections were applied using TWINABS~\cite{twinabs}. No such twinning was observed from the data for [Cu(pyz)$_2$(pyO)$_2$](PF$_6$)$_2$. Each cell component was refined using SAINTPlus~\cite{saintplus} for all observed reflections. Data reduction and correction for Lp and decay were performed. The structure was found by direct methods and refined by a least squares method on F2 using the SHELXTL program suite~\cite{shelxt}. The structures  were solved in space groups $P2_1/m$ (\#11) and $Cmca$ (\#64) by analysis of systematic absences. All non-hydrogen atoms were refined anisotropically. No decomposition was observed during data collection. 

The lattice parameters so determined are shown in Table~\ref{tablestruc}. Also shown are the distances between Cu ions along the crystal axes. Alternate tilting of the pyz and pyO rings along certain directions, as well as staggering of Cu\textemdash pyz planes (see Figs.~1 and 2), leads to unit cells that contain more than one formula unit.
Table~\ref{tablecoord} contains the coordination sphere bond lengths for both materials. The atom numbering schemes can be found in Supplemental Figures S1 and S2, which respectively show the thermal ellipsoid plots for the planar and chain-like materials.

Crystallographic data (.cif format) for the materials described here will be uploaded to the Cambridge
Crystallographic Data Centre.

\vspace{0.2cm}
\noindent{\bf Muon-spin relaxation.} Zero field muon-spin relaxation ($\mu^{+}$SR) measurements were made using the LTF instrument at the Swiss Muon Source at the Paul Scherrer Institute. Powder samples were mounted in vacuum grease on a silver plate, which was fixed to the cold-finger of a 
dilution refrigerator.

In a typical $\mu^{+}$SR experiment \cite{blundell_99} spin-polarized positive muons are stopped in a target sample, where the muon usually
occupies an interstitial position in the crystal. The observed property in the experiment is the time-evolution of the muon-spin polarization, the behaviour of which depends on the
local magnetic field at the muon site. Each muon decays, with an average 
lifetime of 2.2~$\mu$s, into two neutrinos and a positron, the
latter particle being emitted preferentially along the instantaneous direction of the muon spin.
Recording the time dependence of the positron emission directions therefore allows the determination of the spin-polarization of the ensemble of muons. In our experiments positrons are
detected by detectors placed forward (F) and backward (B) of the initial muon polarization direction.
Histograms $N_{\mathrm{F}}(t)$ and $N_{\mathrm{B}}(t)$ record the number of positrons detected in the two detectors as a function of time following the muon implantation. The quantity of interest is
the decay positron asymmetry function, defined as
\begin{equation}
 \renewcommand\theequation{S1}
A(t)=\frac{N_{\mathrm{F}}(t)-\alpha_{\mathrm{exp}} N_{\mathrm{B}}(t)}
{N_{\mathrm{F}}(t)+\alpha_{\mathrm{exp}} N_{\mathrm{B}}(t)} \, ,
\end{equation}
where $\alpha_{\mathrm{exp}}$ is an
experimental calibration constant. $A(t)$ is proportional to the
spin polarization of the muon ensemble, and as a result depends upon the local magnetic field at the muon sites. 

As described in the main text, the planar sample [Cu(pyz)$_{2}$(pyO)$_{2}$](PF$_{6}$)$_{2}$ exhibits oscillations in the $\mu^{+}$SR asymmetry below about 1.7~K, strongly indicative of long range magnetic order throughout the bulk of the material. This is because the local field causes a coherent precession of the spins of those muons for which a component of their spin polarization lies perpendicular to this local field (expected to be 2/3 of the total spin polarization for a powder sample).  The frequency of the oscillations is given by $\nu_{i} = \gamma_{\mu} B_{i}/2 \pi$, where $\gamma_{\mu}$ is the muon gyromagnetic ratio ($=2 \pi \times 135.5$~MHz T$^{-1}$) and $B_{i}$ is the average magnitude of the local magnetic field at the $i$th muon site. The precession frequencies are proportional to the magnetic order parameter. The evolution of the muon precession frequency is shown in the inset to Fig.~2A of the main text and fitting this to the phenomenological function  $\nu(T) = \nu(0)\left[1-(T/T_{\mathrm{N}})^{\alpha} \right]^{\beta}$ yields a transition temperature of  $T_{\mathrm{N}}=1.71\pm0.02$~K and exponents $\alpha = 1.1\pm0.3$ and $\beta = 0.22\pm0.02$. The analysis of $\mu^{+}$SR data for planar coordination polymers is described fully in Ref~\cite{steele11}. 

The $\mu^{+}$SR data for the chain-like compound  [Cu(pyz)(pyO)$_{2}$(H$_{2}$O)$_{2}$](PF$_{6}$)$_{2}$ exhibit no resolvable oscillations, but a monotonic relaxation of the muon polarization across the measured temperature regime (main text, Fig.2B). As previously mentioned, the data were fitted to the expression $A(t) = A_{0}e^{-\lambda t} + A_{\parallel}$, 
where $A_{\parallel}$ represents the non-relaxing part of the signal. Around a 
temperature of $T=0.26$~K we observe a large change in the behaviour of both $A_{\parallel}$ and $\lambda$ which we attribute to the onset of  long-range magnetic order below that temperature. The increase seen in $A_{\parallel}$ on cooling is characteristic of a transition from a region of dynamic magnetic fluctuations to one of quasi-static magnetic order with decreasing temperature~\cite{dalmas_97}. This is attributable to the fact that above $T_{\mathrm{N}}$ dynamic field fluctuations will relax all muon spins; but upon ordering in a polycrystalline material, approximately 1/3 of the muon spins will be aligned along the direction of the quasi-static local field and therefore will not be relaxed. These spins make a non-relaxing contribution to the spectra, detected as an increase in $A_{\parallel}$~\cite{dalmas_97}. In addition, the relaxation rate $\lambda$ is expected 
to vary as a simple function of the second moment of the magnetic field distribution $\langle B^{2} \rangle$ and its sudden increase below $T_{\mathrm{N}}$ reflects the increase in the size of the ordered moment (and hence $\langle B^{2} \rangle$) with decreasing $T$, below the transition. The onset of both effects occurs at the same temperature, from which we estimate the transition temperature of $T_{\mathrm{N}} = 0.27\pm0.01$~K. 

\vspace{0.2cm}
\noindent{\bf Hamiltonian and exchange constants.} Our Hamiltonian (main text, Equation 1) employs the single-$J$ convention such that, in the case of just two $S=1/2$ interacting spins, the separation between the singlet and triplet energies is equal to the interaction strength $J$. In some previous works, the Heisenberg model has been defined such that the singlet-triplet gap is $2J$~\cite{bonnerfisher,woodward_07}. The exchange couplings derived from this definition are half those obtained by our model. Note also that, for materials with two different interchain exchange paths, $J_\perp$ in this Hamiltonian represents an effective interchain exchange.

\vspace{0.2cm}
\noindent{\bf Pulsed-field magnetization.}  Measurements up to 60~T (rise-time to full field $\sim10$~ms) were performed  at the National High Magnetic Field Laboratory in Los Alamos. Single crystals were mounted in 1.3~mm diameter PCTFE ampoules that can be moved into and out of a 1500-turn, 1.5~mm bore, 1.5~mm long compensated-coil susceptometer, constructed from 25~$\mu$m high-purity copper wire. When the sample is within the coil and the field pulsed the voltage induced in the coil is proportional to the rate of change of magnetization with time. Accurate values of the magnetization are then obtained by subtraction of the signal from that taken using an empty coil under the same conditions, followed by numerical integration. The magnetic field is measured via the signal induced an adjacent empty 10-turn coil and calibrated via observation of de Haas--van Alphen oscillations arising from the copper coils of the susceptometer. The susceptometer is placed inside a $^3$He cryostat, which can attain temperatures as low as 0.5~K. Several orientations of single-crystal samples were used. The differences between different orientations of the same sample were found to be small, and in keeping with the $g$-factor anisotropy determined by electron-spin resonance. The data shown in the paper were taken with ${\bf B}\parallel ab$-plane and $T=1.5$~K for the planar material, and ${\bf B}\parallel{\bf b}$ and $T=0.5$~K for the chain-like material. 

The magnitude of $J$ extracted from the saturation field (taken to be the mid-point of the transition in ${\rm d}M/{\rm d}B$) explains why the pulsed-field data for the chain-like material is reminiscent of a Q1D antiferromagnet even at temperatures above $T_{\rm N}$. The short-range correlations that occur above the transition give rise to a hump in both the susceptibility and heat capacity~\cite{johnston_00} of a Q1D Heisenberg antiferromagnet at temperatures $T\sim J$. The temperature at which the pulsed-field data were taken ($T=0.5$~K) is considerably lower than this, and much closer to $T_{\rm N}=0.27$~K, implying that the correlations have grown to such an extent that for a non-local probe it is hard to distinguish this state from one with true long-range order.

\begin{table}[t]
\centering
 \renewcommand\thetable{S3}
\begin{tabular}{|l|c|c|r|}
\hline
 & Q2D & Q1D & technique\\
\hline
$T_{\rm N}$ (K) & $1.71\pm0.02$ & $0.27\pm0.01$ & $\mu^{+}$SR\\
$B_{\rm c}$ (T) & $23.7\pm0.8~~$ & $12.8\pm0.4~~$ & $M(B)$\\
$J$ (K) & $8.1\pm0.3$ & $8.8\pm0.2$ & $M(B)$\\
$J$ (K) & $8.10\pm0.01$ & $8.58\pm0.01$ & $\chi(T)$\\
$g_{xy}$ & $2.04\pm0.01$ & $2.06\pm0.01$ & EPR\\
$g_z$ & $2.26\pm0.01$ & $2.32\pm0.01$ & EPR\\
$J_\perp/J$ & $1\times10^{-4}$ & $1\times10^{-2}$ & QMC\\
\hline
\end{tabular}
\caption{Magnetic properties of [Cu(pyz)$_2$(pyO)$_2$](PF$_6$)$_2$ (Q2D) and [Cu(pyz)(pyO)$_2$(H$_2$O)$_2$](PF$_6$)$_2$ (Q1D) determined using $\mu^{+}$SR -- muon-spin rotation, $M(B)$ -- pulsed-field magnetization, $\chi(T)$ -- low-field magnetic susceptibility, ESR -- electron-spin resonance, and QMC -- quantum Monte Carlo calculations. For the Q2D (Q1D) material the $g$-factors $g_{xy}$ and $g_z$ are measured with the magnetic field within the planes (chains), and parallel to the ${\bf c}$-axis (perpendicular to the Cu\textemdash pyO bond), respectively.}
\label{tablemag}
\end{table}

\vspace{0.2cm}
\noindent{\bf Quantum Monte Carlo simulations.} The numerical results were obtained using the Stochastic Series Expansion (SSE) method to simulate the Hamiltonian of Equation~1 on finite-sized lattices. SSE is a finite temperature QMC method based on the evaluation of the diagonal matrix elements of the density matrix in a suitable basis. The method is explained in detail in Ref.~\cite{sengupta_03} and references therein. The simulations were performed on the NERSC computing facility in Berkeley, California.

\begin{figure}[t]
\centering
 \renewcommand\thefigure{S3}
\includegraphics[width=8cm]{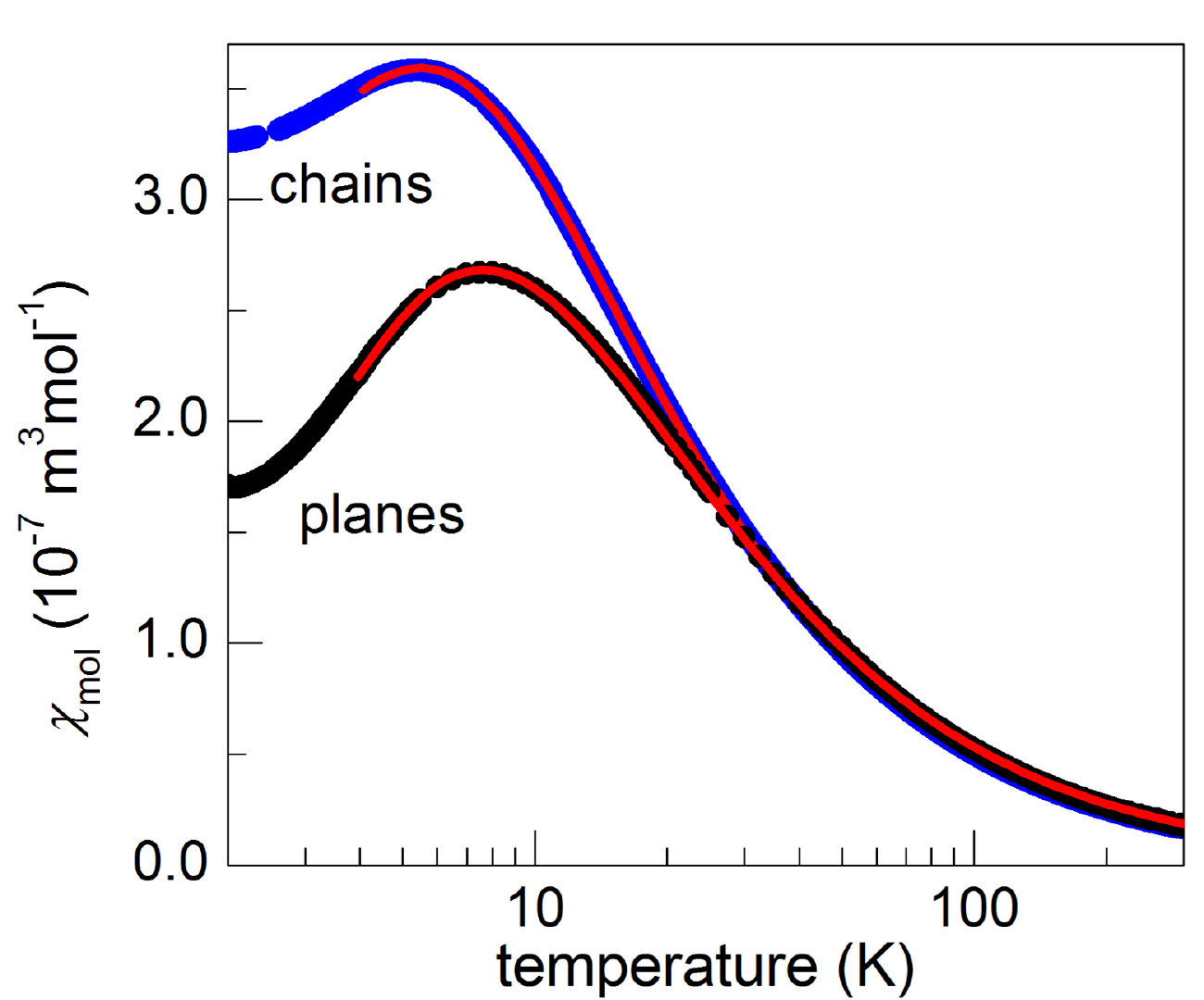}
\caption{Results of temperature-dependent susceptibility measurements. Data points are the powder magnetic susceptibilities for both materials, red lines are fits to low-dimensional models described in the text.} 
\label{chi}
\end{figure}

\vspace{0.2cm}
\noindent{\bf Temperature-dependent susceptibility.} Measurements were performed using a Quantum Design Physical Properties Measurement System equipped with a vibrating sample magnetometer option. Polycrystalline samples were placed into gelatin capsules and affixed to the end of a carbon fibre rod.  Data were taken on warming from 2~K in a fixed magnetic field of 0.1~T. 
The datasets for the two compounds are quite distinct (see Fig.~\ref{chi}), but both exhibit the broad hump in ${\sl \chi}_{\rm mol}(T)$ at $T\sim J$ expected for highly anisotropic antiferromagnets and caused by the onset of short-range correlations. Fits from 300~K down to temperatures just below the hump were made to theoretical models based on the low-dimensional Heisenberg Hamiltonian of Equation~1. The layered compound is found to be in excellent agreement with the $S=1/2$ Q2D model of Woodward {\it et al.}~\cite{woodward_07} with $g=2.21$ and $J=8.10\pm0.01$~K, while the best fit for the chain-like material is for the quasi-one-dimensional model of Johnston {\it et al.}~\cite{johnston_00} with $g=2.11$ and $J=8.58\pm0.01$~K. These values are in good agreement with those derived from the pulsed field magnetization. We obtained similar results by also fitting to the earlier models of Bonner \& Fisher~\cite{bonnerfisher} (Q1D) and Lines~\cite{lines} (Q2D). 

\vspace{0.2cm}
\noindent{\bf Electron-spin resonance.} Measurements of the $g$-factor anisotropy were performed in a superconducting magnet at the National High Magnetic Field Laboratory in Los Alamos using a cavity perturbation technique~\cite{mcdonald_06}. The TE102 mode of a rectangular resonator at 71.5 GHz, and a Millimetre-wave Vector Network Analyser manufactured by ABmm were used. The cavity was mounted on a goniometer and placed inside standard flow $^4$He and single-shot $^3$He cryostats. A temperature dependence (down to 0.5 K) of the spectrum along the principal axes was performed to ensure the reported angular dependance of the resonance is not influenced by the long range order, but characteristic of the paramagnetic state. For both compounds the data shown satisfy this condition. 

\vspace{0.2cm}
The magnetic properties of the two materials studied are summarized in Table~\ref{tablemag}.

\end{document}